\documentclass[10pt, epsfig]{article}                            
\usepackage{epsfig}       
\usepackage{amssymb}     
\usepackage{amsfonts}    
\newskip\humongous \humongous=0pt plus 1000pt minus 1000pt

\newif\ifdtup

\catcode`\@=11
 
\@addtoreset{equation}{section}
\def\theequation{\thesection\arabic{equation}}
 
\def\@normalsize{\@setsize\normalsize{15pt}\xiipt\@xiipt
\abovedisplayskip 14pt plus3pt minus3pt%
\belowdisplayskip \abovedisplayskip
\abovedisplayshortskip \z@ plus3pt%
\belowdisplayshortskip 7pt plus3.5pt minus0pt}
 
\def\small{\@setsize\small{13.6pt}\xipt\@xipt
\abovedisplayskip 13pt plus3pt minus3pt%
\belowdisplayskip \abovedisplayskip
\abovedisplayshortskip \z@ plus3pt%
\belowdisplayshortskip 7pt plus3.5pt minus0pt
\def\@listi{\parsep 4.5pt plus 2pt minus 1pt
     \itemsep \parsep
     \topsep 9pt plus 3pt minus 3pt}}
 
\relax

\catcode`@=12
 
\catcode`\@=11
 
\def\section{\@startsection{section}{1}{\z@}{3.5ex plus 1ex minus
   .2ex}{2.3ex plus .2ex}{\large\bf}}
 
\def\thesection{\arabic{section}.}

\def\appendix{\setcounter{section}{0}
 \def\thesection{Appendix \Alph{section}:}
 \def\theequation{\Alph{section}.\arabic{equation}}}
\def\YGrule{0.4}   % line thickness in unit of pt
\def\YGbox{6.5}    % box size in unit of pt
\def\SymBoxes#1#2#3#4{\newdimen\un@t \un@t#3%
\raisebox{#1}{\rule{#2\un@t}{#4}\hskip-#2\un@t% lower horizontal
\@tempdimb\un@t \advance\@tempdimb by-#4\@tempcntb#2\relax%
\@whilenum{\@tempcntb>0}\do{%                         % #2 vertical lines
\rule{#4}{\un@t}\hskip\@tempdimb \advance\@tempcntb by\m@ne}%
\hskip-#2\un@t \rule[\un@t]{#2\un@t}{#4}%
\rule[\un@t]{#4}{#4}\hskip-#4%             % upper horizontal line
\rule{#4}{\un@t}}\hskip-#4}                % rightest vertical line
\def\Young{\@ifnextchar[{\@Young}{\@Young[0]}}
\def\@Young[#1]#2{\newdimen\YG@unit \YG@unit\YGbox pt%
\newdimen\h@ight \h@ight#1\YG@unit \@tempcnta-1\relax
\@tfor\c@ount:=#2\do{\advance\@tempcnta by\@ne}% count the number of rows
\@tempdima\@tempcnta\YG@unit%
\advance\h@ight by\@tempdima\relax     % compute the height of the top row
\@tfor\c@ount:=#2\do{\SymBoxes{\h@ight}{\c@ount}{\YG@unit}{\YGrule pt}%
\@tempdima-\c@ount\YG@unit \hskip\@tempdima%
\advance \h@ight by -\YG@unit}         % Draw the Tableaux
\@tempdima\YG@unit \multiply\@tempdima by\@car#2\@nil %
\hskip\@tempdima}                      % hskip by the length of the top row
\def\YoungTab{\@ifnextchar[{\@YoungIdx}{\@YoungIdx[0]}}
\def\@YoungIdx[#1]{\@ifnextchar[{\@iYoungIdx[#1]}{\@iYoungIdx[#1][\@empty]}}
\def\@iYoungIdx[#1][#2]#3{%
\newdimen\YG@unit \YG@unit\YGbox pt\newdimen\YG@rule \YG@rule \YGrule pt
\newcount\c@ount \c@ount\z@ \newdimen\skip@wd \unitlength\@ne pt
\newdimen\h@ight \h@ight#1\YG@unit \@tempcnta\m@ne\relax
\@tfor\d@um:=#3\do{\advance\@tempcnta by\@ne}% count the number of rows
\@tempdima\@tempcnta\YG@unit%
\advance\h@ight by\@tempdima\relax%  % compute the height of the top row
\@tfor\@idxlist:=#3\do{%             % routine to draw the indexed Tableaux
\@tempcnta\z@\hskip.5\YG@rule\relax 
\@for\@idx:=\@idxlist\do{%           % place the indices of the row first
\raisebox{\h@ight}{\makebox(\YGbox,\YGbox){#2$\@idx$}}
\advance\@tempcnta by\@ne}\hskip-.5\YG@rule% 
\@tempdima-\@tempcnta\YG@unit \hskip\@tempdima%
\ifnum\c@ount=\z@ \skip@wd-\@tempdima\fi \relax% record the top row width
\SymBoxes{\h@ight}{\@tempcnta}{\YG@unit}{\YG@rule}%
\hskip\@tempdima \advance\h@ight by -\YG@unit
\advance\c@ount by\@ne}%             % end of the routine
\hskip\skip@wd}                      % hskip by the length of the top row
\begin{document}
%\begin{letter}{~}

%%%%%%Define some new commands and  macros
\newcommand{\beq}{\begin{equation}}
\newcommand{\eeq}{\end{equation}}
\newcommand{\bea}{\begin{eqnarray}}
\newcommand{\eea}{\end{eqnarray}}
\newcommand{\beas}{\begin{eqnarray*}}
\newcommand{\eeas}{\end{eqnarray*}}
\newcommand{\defi}{\stackrel{\rm def}{=}}
\newcommand{\non}{\nonumber}   
\newcommand{\bquo}{\begin{quote}}
\newcommand{\enqu}{\end{quote}}
%%%%%%%%%%%%%%%%%%%%%%%%%%%%%%%%%% definitions
\def\de{\partial}
\def\Tr{ \hbox{\rm Tr}}
\def\const{\hbox {\rm const.}}
\def\o{\over}
\def\im{\hbox{\rm Im}}
\def\re{\hbox{\rm Re}}  
\def\bra{\langle}\def\ket{\rangle}
\def\Arg{\hbox {\rm Arg}}
\def\Re{\hbox {\rm Re}}
\def\Im{\hbox {\rm Im}}
\def\diag{\hbox{\rm diag}}
\def\longvert{{\rule[-2mm]{0.1mm}{7mm}}\,}

\begin{titlepage}
{\hfill     IFUP-TH 20/2001} 
\bigskip
\bigskip
\bigskip
\bigskip

\begin{center}
{\large  {\bf 
  NON-ABELIAN VORTEX      AND CONFINEMENT  
} }
\end{center}

\vspace{1em}

\begin{center}
{ KENICHI  KONISHI    }
\end{center}     

\begin{center}
{\it
Dipartimento di Fisica   ``E. Fermi"  -- Universit\`a di Pisa\\
Istituto Nazionale di Fisica Nucleare -- Sezione di Pisa  }
\\
{\it Via Buonarroti, 2,   Ed. C, 56127  Pisa, Italy}  \\
\end{center} 

\vspace{1em}

\begin{center}
{ LEONARDO SPANU   }
\end{center}     

\begin{center}

{\it
 Dipartimento di Fisica   ``E. Fermi"  -- Universit\`a di Pisa }
\\
{\it Via Buonarroti, 2,   Ed. C, 56127  Pisa, Italy}  \\  

\end {center}

\vspace{1em}
   
{ We discuss      general   properties and  possible types  of magnetic  vortices  in   non-Abelian
gauge theories   (we consider here  $G= SU(N), SO(N), USp(2N)$)   in the  Higgs    phase.    The sources of such vortices     carry  ``fractional"
quantum numbers such as
$Z_n$ charge (for $SU(N)$),    but also full non-Abelian charges of the dual gauge group.      If  such a model  emerges   as  an effective dual
magnetic  theory   of     the  fundamental (electric)  theory,     the non-Abelian vortices can  provide for the mechanism of
quark-confinement in the latter.     }

\vfill
 
\begin{flushright}
June 2002
\end{flushright}
\end{titlepage}

\bigskip

\section{Introduction} 

The mechanism of confinement and dynamical symmetry breaking,  and the relation theirof,   has recently been  studied in detail,   
in a class of asymptotically  free   ${\cal N}=2$ supersymmetric gauge theories (in which  supersymmetry is sofly broken to  ${\cal
N}=1$) with various gauge groups,   $SU(n_c), USp(2n_c)$ and  $ SO(n_c)$,      and with different
numbers of flavors \cite{CKMP},    generalizing the pioneering  works of Seiberg and Witten \cite{SW1,SW2}.    These models  are
characterized by the existence of a large (discrete)    vacuum degeneracy,  even in  the presence of
nonvanishing, generic matter   masses,  and  as a result  a theory with a given Lagrangean  can and does
   in  fact  display distinct  vacua in
various  phases. 

An  advantage  of studying such theories lies in the fact that the low-energy degrees of freedom and the 
form of the effective action can be determined explicitly,  by combining duality,   supersymmetry, Seiberg-Witten
exact curves \cite{curves}   and  some knowledge on  superconformal theories \cite{SCF}.   Thus we learn that, 
even if we  restrict ourselves to    vacua in confinement phase,   there are   distinct  types  of confining 
phases,   distinguished by different  entities  which condense. 
In some vacua   they are monopoles of the maximal Abelian subgroup of the gauge group $G$,  as envisaged by 't Hooft and 
Mandelstam \cite{TM}.   More typically, however,   they are  magnetic monopoles (dual quarks)  carrying  non-Abelian charges,   and
interacting  with light gauge bosons of a non-Abelian  effective gauge theory.    It also happens  that,  as in an important class of vacua in
$SU(n_c)$, $USp(2n_c)$  and 
$SO(n_c)$ theories,   confinement is due to  the cooperation of  relatively   non-local dyons.     The effective theory is near 
 a nontrivial superconformal infrared fixed point   in these cases. 

Also,   the question  of which   fields appear as the low-energy effective degrees of freedom,  was
found to be intimately  related     to the pattern of dynamical symmetry breaking \cite{CKMP}.

Another interesting model is the $SU(N)$ gauge theory with $N=4$  supersymmetry, in which supersymmetry is softly broken to $N=1$
by  the  masses of the three adjoint scalar multiplets.   Dual Meissner effect occurs in the confining   vacuum of this model, but with
no sign of   dynamical Abelianization \cite{STRASS}.

In the case of  the   standard, non-supersymmetric QCD,   where there is a unique vacuum,   the system  must 
choose a particular type of confinement mechanism.  
  Which among the above mentioned possibilities is realized  in QCD is not yet known. 
Certain problems in the Abelian mechanism of confinement for QCD, have been pointed out \cite{DS,STRASS,YUNG}.

Inspired  by  these new developments,   we propose here to give a renewed look into   the  general    properties of vortices appearing in
a  non-Abelian gauge theory,  in which the continuous gauge symmetry is completely  broken 
by the Higgs mechanism,  leaving  a discrete center unbroken.     These vortex solutions are generalizations of 
Nielsen-Olesen-Abrikosov  vortices  of   $U(1)$  gauge theory \cite{ABR,NO},    but  possess   ``fractional" 
 quantum numbers,   such  as     $\mathbb Z_N$   charge  ($N$-ality),   for instance,   in the case of $SU(N)$ theory.  
Quite detailed studies  of explicit vortex-like solutions  in the cases of $SU(N)$ theories   have been made    earlier by de Vega
and  Schaposnik \cite{DeVega,DS1} and more
recently  by several authors 
\cite{HV}-\cite{Kneipp} by using     some explicit  models  of  scalar potentials.    
See also   \cite{Hase, Edel}.

In spite of these efforts, we feel that a systematic study of non-Abelian vortices in general is still lacking.   We therefore present
the following analysis  as a first step towards that end,  even though   we  are  fully    aware  that some  overlap with  earlier 
results  is inevitable.  The main interests here will be some general properties of the vortices such as  the   quantum
numbers carried by their possible sources.    
Study of minimul vortex solutions  in  various  types of  gauge theories  nicely shows how the 
the quantum numbers  of the  
 dual gauge groups introduced by  Goddard, Olive and Nuyts \cite{GNO}  make  natural appearance associated with the  sources of these
vortices.

For 
instance,     the $G=SU(N)/{\mathbb Z}_N$   theory  ($SU(N)$ theory  without fundamental matter),   has quantized  vortices  
which can terminate at sources  in  a fundamental representation   of 
 ${\tilde G}  =
SU(N)$ group  (with  ${\mathbb Z}_N$ charge).  More general solutions (not all stable)  are associated with  sources  corresponding
 to
irreducible multiplets    of 
${\tilde G}$,  represented     by  the Young    tableaux with
$n=1,2,\ldots N-1$  boxes.

Which of these vortices other than  the 
lowest, $N$-ality one  vortex,   represent  stable vortices (and not just  bundles of lower  vortices),    
and what   the relative tensions  among them  are, etc.,    are   questions
depending on the details of dynamics,  quantum effects, etc.,   going   beyond  the scope of the 
 general discussions in this   paper.

\section{General  Characterization of Vortices: Flux Quantization } 

We consider a gauge theory  in which the gauge group $G$ is spontaneously broken by the Higgs mechanism as 
\beq   G \Longrightarrow     {\cal C} 
\eeq
with $  {\cal C} $   a discrete center of the group.   The general properties of the vortex, which represents a nontrivial elements of the 
fundamental group,
\beq    \Pi_1(G/C)  = C, 
\eeq
    are  
independent of the detailed form of the scalar potential or of the number of  the Higgs fields present: they are determined by the  
asymptotic behavior of the fields.  The latter   should be such that far from the vortex   the gauge fields are  pure gauge form, and the 
matter scalar   fields are covariantly constant and at the minimum of the potential. 
With  an appropriate gauge choice such fields can be taken, far from the core of the vortex (which we take along the $z$ axis),      
as
\beq   A_i  \sim   { i \o  g}    \, U(\phi) \de_i  U^{\dagger}(\phi); 
\quad     \phi_A \sim   U \phi_{A}^{(0)}  U^{\dagger},     \qquad  U(\phi) = \exp{i \sum_j^r 
{\bf {\beta}}_j T_j  
\phi}
\label{standard} \eeq
where  $\phi_A^{(0)}$  are  fixed  scalar VEVS  at a minimum of the potential,  and     $T_i$'s
are  the generators of the Cartan subalgebra of $G$.  Since $T_i$'s commute with each other,  one has  
\beq   A_{\phi}  \sim   { 1 \o  g  \, r  }  \sum_j^r  \beta_j T_j 
\eeq
so that the above vortex   carry   the flux
\beq  \oint  dx_i A_i    =  { 2 \pi  \o g   }   \sum_j^r  \beta_j T_j.
\eeq
Such a flux is well-defined modulo  gauge transformation which permute $T_i$'s  (Weyl  transformations), hence  
$\beta_i$'s: \footnote{We follow the notation of Goddard, Nuyts and Olive  \cite{GNO}.  }   
\beq   S_{\alpha}   \left(  \sum_j^r  \beta_j T_j  \right)    S_{\alpha}  =  \sum_j^r  \beta_j^{\prime}  T_j, \qquad 
  \beta^{\prime}=  {\bf \beta } - { 2 {\bf \alpha}  ( \bf {\beta  }   \cdot {\bf \alpha}) \o  ( {\bf \alpha} \cdot  {\bf
\alpha})},
\label{weyltr} \eeq
\beq      S_{\alpha}  =  \exp[ i \pi (E_{\alpha} + E_{-\alpha}) / \sqrt  { 2 \alpha^2}  ], 
\eeq 
where  $ \alpha$ is a root vector   and $E_{\alpha}$  is a nondiagonal generator in the     Cartan basis.

 The quantization condition on  $\beta_j$  follows from the requirement that the    fields      are single valued,      i.e.,
\beq  U(2 \pi)  \in    {\cal C}. 
\label{qcondi1}  \eeq
 For $SU(N)$ with all  scalar fields in the adjoint representation,   ${\cal C}= \mathbb Z_N$.    For $SO(2N)$, $N \ge 2$  with all  scalar 
fields in
the vector  representation, 
 ${\cal C}= \mathbb Z_2$.  
The fact that  $U(2 \pi)  \in    {\cal C}$  commutes with all generators leads, by commuting it with $E_{\alpha}$'s,
to  the general codition  for $\beta$, \footnote{    Note that our $\beta$   are twice as large as
compared to
$\beta$ appearing in the problem of non-Abelian  monopoles \cite{GNO}. }  
\beq  {\bf \alpha} \cdot { \bf \beta} \in  \mathbb{Z}.       
\label{general} \eeq

Thus the minimum vortex of $SU(N)$, for instance,    corresponds to   $U_{min} (\phi)$  with 
\beq   U_{min} (2 \pi) =   e^{2 \pi i / N}  \,  {\bf 1},   
\eeq 
which represents  a noncontractible loop of  the coset space $SU(N)/\mathbb Z_N$.    
On the other hand,    a bundle of $N$ vortices or a single vortex with $N$ windings
\beq  U(\phi) = \exp{i N  \sum_j^r   \beta_j T_j
\phi} = [U_{min} (\phi)]^N
\eeq
satisfies  
\beq   U(2 \pi)={\bf 1}:
\eeq
it is a loop in  $SU(N)$.   Since $SU(N)$ is simply connected,    such a loop can be  smoothly  
contracted to a point,   as $r$ is reduced from    $r= \infty$    to $r=0$.

The general  condition Eq.(\ref{general})  implies that the ``charges" $\beta$ characterizing   the vortex  live in the
dual   group  ${\tilde  G}$    of $G$,   as shown   by   Goddard, Nuyts and Olive \cite{GNO}  in an  analogous problem with  the  monopoles.   
This follows from the fact  that  the  root vectors of the  group in which  the charges  $\beta$    live  form  the dual lattice of    the original
root lattice.    The examples of such dual  groups are:
\bea   SU(N) & \leftrightarrow   &   SU(N)/\mathbb Z_N;  \non \\
 SO(2N) & \leftrightarrow   &   SO(2N);  \non \\
 SO(2N+1) & \leftrightarrow   &   USp(2N).   
 \label{dualgroup}   \eea

Physical settings studied by Goddard, Nuyts and Olive \cite{GNO}  (monopoles)  are slightly  different from the ones studied here
(vortices).    Regular  monopole solutions  appear in   a  theory  in which the   gauge group    $G$ is  broken spontaneously to an unbroken 
subgroup $H$ by Higgs mechanism, 
\beq    G   \,\,\,{\stackrel {\bra \Phi \ket    \ne 0} {\longrightarrow}}    \,\,\, H. \eeq 
such that 
\beq    \Pi_2(G/H)  \ne {\bf 1}. 
\eeq  
($\Pi_2(G/H)  = \Pi_1(H)$ if   $G$ is simply connected).  
The first example of such monopole is nothing but the 't Hooft-Polyakov monopole of   
$G = SU(2)$,  $\,H=U(1)$  theory  \cite{TOPO}.       Their possible   charges,  with respect
to the  (generally, non-Abelian)  unbroken  group $H$,   are  characterized      by  the   weight vectors of the dual group  ${\tilde 
H}$  \cite{GNO}.

\section  { An Illustration:   $SO(3)=SU(2)/\mathbb Z_2$   Vortex  \label{sec: so3}}
 
For the purpose of making the paper self-contained,    we first briefly illustrate  the simplest non-trivial case 
of a $Z_2$   vortex.   
Consider   a $SO(3)=SU(2)/\mathbb Z_2$  model with   the Lagrangean, 
  \beq
 L=   - { 1\o 4}  {\hat F}_{\mu \nu}^2  +  { 1\o 2}  [(D_{\mu} {\hat  \phi_1 })^2  +  (D_{\mu} {\hat \phi_2 } )^2]
 -   V({\hat  \phi_1 }, {\hat \phi_2 }),
\eeq
  where 
 \beq {\hat F}_{\mu \nu} =  \de_{\mu}  {\hat A}_{\nu} -  \de_{\nu}  {\hat A}_{\mu} + g   {\hat A}_{\mu} \times {\hat A}_{\nu}; 
\qquad      D_{\mu} {\hat  \phi_i } =   (\de_{\mu}  +  g {\hat A}_{\mu} \times ) {\hat  \phi_i };  
\eeq
and the potential can be taken,  for instance,    as 
\beq     V({\hat  \phi_1 }, {\hat \phi_2 })=   \sum_{A=1}^2   \lambda_A ( {\hat  \phi_A } \cdot {\hat  \phi_A } - F_A^2)^2  +
\kappa   ( {\hat  \phi_1} \cdot {\hat  \phi_2 } -G^2)^2. \eeq
The potential is taken  such that  the gauge group  $SO(3)$ is  broken spontaneously      at its minima.  
The minimum of the potential is    at    
\beq    {\hat \phi_A } \cdot   {\hat \phi_A }  =F_A^2  \quad ( A=1,2 );   \qquad 
   {\hat \phi_1 }\cdot   {\hat \phi_2 }   ={G^2 } \equiv F_1 F_2  \cos \Theta, \eeq 
where we have    assumed  $ | {G^2 \o F_1 F_2}|<1.$      By a gauge transformation  ${\hat \phi_1 }$  can be taken in the  $3$ direction
  (of the isospin space). 
Existence of regular  solutions of  vortex type  can be easily worked out,    which have    the 
asymptotic behavior   at  large  $r$  (where  we introduce the
cylindrical  coordinates $(r, \phi, z)$)  
\beq   {\hat \phi_1}  \sim   U_n(\phi)   \pmatrix{F_1 \cr 0 \cr 0}; \qquad    
 {\hat \phi_2}  \sim    U_n(\phi)  U_1(\Theta ) \pmatrix{ F_2   \cr 0 \cr 0};    \qquad  A_i  \sim   U_n(\phi)  {1 \o g}  
\de_i     U_n(\phi) ^{\dagger}, 
\label{candidate} \eeq
with 
\beq      U_n(\phi) =  \pmatrix {\cos  n  \phi  &  -  \sin n  \phi    & 0   \cr     \sin n \phi  & \cos n \phi   & 0 \cr 0 & 0 & 1}. 
\label{asympgauge}   \eeq 
As this analysis  overlaps  substantially    with   the results of earlier
works \cite{DeVega}-\cite{SS}, we restrict ourselves to a  brief discussion given  in Appendix A. 
Note that ${\hat A}_{i}$ and  ${\hat A}_{ij} $ are nonvanishing only for   $i=1,2$.
As one goes around the string (which is taken to be along the $z$ axis),  the vector   ${\hat \phi_1, }$ ${\hat \phi_2 }$ 
rotate   $n$ times around the   $3$ axis of the $SO(3)$ space: it  gives  rise apparently  to a vortex with  winding number $n$.

Indeed,  following   \cite{NO}  one can show that for a   static vortex,    
\beq      \int_{S_1} {\hat F}_{ij} \, d\sigma_{ij}  -   \int_{S_2} {\hat F}_{ij} \,d\sigma_{ij}  =0,  \qquad  i,j=1,2
\eeq 
where  $S_{1,2}$  are the  bottom and    top circles   of the cylindrical region considered, 
 apparently  allowing  one  to  define a ``conserved"   flux.   For the above mentioned string configuration,  one finds
\beq       \int_{S} { F}_{ij}^3 \, d\sigma_{ij} =   \oint dx_i    { A}_{i}^3  =   { 2 \pi n \o g}, 
\label{winding}  \eeq 
where the circulation is taken at a large radius $r$.

The problem  with this definition of the flux is that it is not gauge invariant.  
In fact, in the case of the vortex with flux two ($n=2$),    it is possible   to construct explicitly  (Appendix B)   a gauge
transformation, regular everywhere,   such that 
\beq    U_{global}(\phi, r)  \stackrel { r \to \infty} {\longrightarrow } U_2( -\phi); 
\qquad       U_{global}(\phi, r)   \stackrel { r \to 0  } {\longrightarrow } {\bf 1}. 
\eeq
By using such a gauge transformation, the apparent vortex can be gauged away.  
   This corresponds to  the well-known fact
the $ 4 \pi $ rotation in the $SO(3)$  space
can be smoothly   shrunk to  a point.  
 On the other hand,  for the minimum winding number $n=1$,  this is not possible due to the fact that
\beq   \Pi_1(SO(3))=    \Pi_1(SU(2)/\mathbb Z_2)  = \mathbb Z_2. 
\eeq
Sources of the  vortices in the Higgs phase of   $SO(3)$   theory   carry   thus  the unique    $\mathbb Z_2$ charge.\footnote{  
For the same mathematical reason, the Dirac like magnetic monopoles in an {\it unbroken}  $SO(3)$    theory  carry the unique
  $\mathbb Z_2$ charge \cite{WUYANG}.}

An analogous procedure might appear  to be capable  of eliminating   the winding of the gauge fields hence the flux analogous to  
(\ref{winding}) in the Abelian,  Abrikosov-Nielsen-Olesen  vortices as well.  The crucial difference is that in the latter case any such gauge
transformation necessarily introduces a singularity in the gauge fields along the core of the vortex ($r=0$):   the flux is now
concentrated within an infinitely thin vortex core:  a vortex cannot be eliminated in an Abelian case, whatever its winding
number   may be. This  fact reflects    the topological property
\beq  \Pi_1(U(1))= \mathbb{Z}.        
\eeq

\section{Quantum Numbers of  the Sources} 

The quantization condition on the vortices Eq.(\ref{qcondi1}), Eq.(\ref{general}) will be solved now explicitly,
and the solutions will be classified,  for   $SU(3)$, $SU(N)$,   $SO(2N)$,  $SO(2N+1)$ and  $USp(2N)$    gauge  groups.

 \subsection{  $\mathbf {SU(3)} $ }
      By normalizing the   generators  in a canonical way,  we have 
\beq   T_3= { 1 \o 2 \sqrt3} \pmatrix {1 & 0 & 0\cr 0 & -1 & 0\cr
0 & 0& 0};\qquad T_8= { 1 \o 6}  \pmatrix {1 & 0 & 0\cr 0 & 1 & 0\cr
0 & 0& -2}.\eeq
The normalization of the generators  is  such that   the metric of the root vector space  satisfies    
\beq    g^{ij} =  \sum_{roots} \alpha^i \alpha^j  = \delta^{ij}.   \label{rootnormal} \eeq
With this normalization   the asymptotin gauge transformation matrix becomes 
\beq    U(\phi) = \exp{i \sum_j^r \beta_j T_j
\phi}  \longrightarrow      \pmatrix  {  e^{  i  \phi  (\beta_3/2 \sqrt{3}  + \beta_8/6  ) } &  0 & 0   \cr 0 
&   e^{  i  \phi  (-\beta_3/2 \sqrt{3}  +\beta_8/6) }  & 0  \cr 
0 & 0 &  e^{  -   i  \phi   \beta_8/3  } }. 
\eeq  
The quantization condition  $  U(2\pi) \in \mathbb Z_3$    gives
\beq  { \beta_3 \o 2\sqrt3 } +  { \beta_8 \o  6} =  - { n_1 \o 3}, \quad   -{ \beta_3 \o 2\sqrt3 } +  { \beta_8 \o  6} =  - { n_2 \o 3},
 \quad
   - { 1\o 3}   \beta_8=  - { n_3 \o 3},   \label{quantiz1}  \eeq 
where  for a  vortex of minimum winding,
\beq      n_i=    {[1 \,\,  mod \,\,    3] },\qquad   \sum_{i} n_i=0.
 \label{quantiz2}  \eeq
The   minimum  solution  for ${\bf \beta}= ( \beta_3,   \beta_8)$   is 
\beq   {\bf \beta}= ( - \sqrt 3, 1 ), \,\,  (  \sqrt 3,  1 ), \,\, {\hbox{\rm or}}    \,\,   (  0, - 2 ), \,\, 
\label{minsol} \eeq
namely,   
\beq   {\bf \beta}=  6 \,  { {\bf w }}  = 2 N   { {\bf w }},   
\label{choicessu3} \eeq  
where ${\ {\bf w }}$  is a   weight vector of the   fundamental representation,   ${\underline {3}}$. 
Obviously, if one chooses  the weight vector   of the anti-fundamental representation,   
\beq   {\bf  \beta}=  6\,   {\bar {\bf w }} = -  6 \,  { {\bf w }},   
\eeq  
one gets instead a vortex with flux  minus   one.

Also, it is clear that by taking vector sums  of the minimum solutions  $3 n +1$  times   ($n=\pm 1, \pm2, \ldots$), 
one constructs an infinite number of (probably all  unstable except for $n=0$)  vortices of triality one.

Various choices for $\beta$  in  Eq.(\ref{choicessu3})   are related by Weyl transformations, 
\beq    {\bf w } \to    {\bf w } - { 2 {\bf \alpha}  ( {\bf w }\cdot {\bf \alpha}) \o  ( {\bf \alpha} \cdot  {\bf \alpha})},
\eeq
which can be   obtained by a gauge transformation of the form, Eq.(\ref{weyltr}).  
In other words, a vortex with one solution of Eq.(\ref{quantiz1}), Eq.(\ref{quantiz2})   and  another vortex 
related to it  by  a Weyl transformation, are gauge equivalent.    
These correspond  precisely to the possible sources  of such  magnetic vortex  which carry  the 
  quantum numbers  of the representation ${\underline {3}}$,  and  a unit triality.       We have a unique   gauge-invariant  
$\mathbb Z_3$  vortex  of triality one.

The solutions with  triality    two  can found by choosing   the solution of 
\beq  { \beta_3 \o 2\sqrt3 } +  { \beta_8 \o  6} =  - { n_1 \o 3}, \quad   -{ \beta_3 \o 2\sqrt3 } +  { \beta_8 \o  6} =   -{ n_2 \o 3},
 \quad
   - { 1\o 3}   \beta_8=  - { n_3 \o 3},   \label{quantiz3}  \eeq 
with   
\beq      n_i=   [2 \,\,  mod \,\,    3],  \qquad   \sum_{i} n_i=0.
 \label{quantiz4}  \eeq
From the point of view of $\mathbb Z_3$ quantum numbers,    of course, the triality two is equivalent to minus   one;    however a priori
 this may not be 
 the   whole   story.    The source of these vortices  can  carry also the full  $SU(3)$  quantum numbers.   In the case of flux two,  there are 
in fact   two solutions,   
\beq   {\bf \beta}= (  -\sqrt 3, -1 ), \,\,  (  \sqrt 3, -1 ), \,\, {\hbox{\rm or}}    \,\,   (  0, 2 ), \,\, 
\label{triplet}   \eeq
which are the weight vectors of   ${\underline  3^{*}}$, and  
   \beq   {\bf \beta}= (  -2 \sqrt 3, 2 ), \,\,  (  2\sqrt 3, 2 ), \,\,   (  0, 2 ).  \,\,    ( - \sqrt 3,  -1 ), \,\,   (  \sqrt 3, -1 ), \,\, 
{\hbox{\rm or}}   
\,\,  (  0, -4 ), \,\,
\label{sextet}   \eeq 
which  are  the weight vectors of    ${\underline  6}$.

\begin{figure}[h]
\begin{center}
\epsfig{file=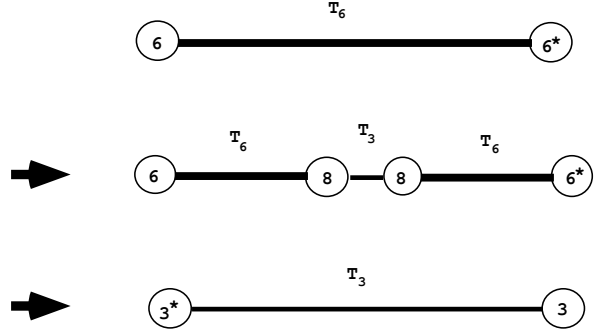,  width=8cm}                
\end{center}
\caption{ A vortex with a  higher tension decays into one with a lower tension through the pair production of
particles in the adjoint   representation. }    
\label{vortex}    
\end{figure}

\begin{figure}[h]
\begin{center}
\epsfig{file=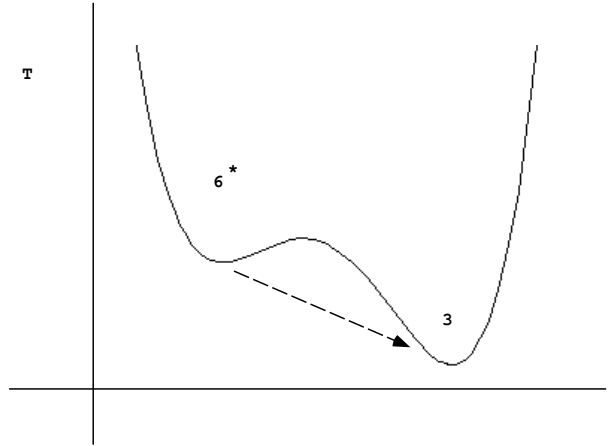,  width=8cm}                
\end{center}
\caption{ The same process as Fig.\ref{vortex}  seen as a tunnel effect in the semiclassical approximation in the Higgs model.}       
\label{vortdecay}    
\end{figure}

Classically these correspond to    two distinct    gauge-invariant sets of vortices.    However, quantum mechanically,   
the vortex with the higher tension   (probably ${\underline  6}$)    will decay   into the
one with the lower tension  (probably ${\underline  3^{*}}$)     through
the  gauge  boson pair productions (Fig. \ref{vortex}, Fig. \ref{vortdecay}) \footnote{Recently a  process  of this sort has been analyzed 
in a slightly different context  by Shifman and Yung \cite{SY}. }.   We thus   find   a  unique $\mathbb Z_3$  vortex in the 
$SU(3)$ gauge theory.

 As   the dual of $SU(3)/\mathbb Z_3$  is precisely $SU(3)$,   the vortex  found above  corresponds to the confining string
for    the  quarks   if the present model is realized  as the magnetic $SU(3)$   theory (see
Sec.\ref{sec:duality}   below). 

Note  that  the same asymptotic   "charges", e.g.,     $\beta_i= (  \sqrt 3, -1 ),$   appear    in the solution for  the  ${\underline  3^{*}}$
  as well as for   ${\underline  6}$.   This shows that  it is not correct in general to study the non-Abelian vortices  by
abelianizing the model (i.e., assuming  nonzero field components only in some Abelian subgroup).  Full non-Abelian equation of motion must
be analysed.

And this leads us  to   another issue which are sometimes overlooked in the literature.  As the vortex  in  $SU(3)/\mathbb Z_3$ theory
carries  $\mathbb Z_3$  flux,   it cannot be BPS  (i.e., the equation cannot be linearized) in general,  and this makes 
the analysis of non-Abelian vortices more difficult than the Abelian case.

 \subsection{    $\mathbf { SU(N)}$   gauge theory }   
The diagonal  generators can be taken as  
\beq
{\bf{T_i} }  =
\left( \begin{array}{ccccc}
w_{1}^{i} & & & & \\
 & w_{2}^{i} & & & \\
 & 0 & \ddots & 0 & \\
 & & & w_{N-1}^{i} & \\  
 & & & & w_{N}^{i}
\end{array} \right) ,   \qquad         i=1,2...N-1
\label{SUNGEN}  \end{equation} 
where   $w_{k}$   represents the $k$-th  weight vector of the fundamental representation 
of $SU(N)$, satisfying \cite{Group}
\beq  {\bf w}_k \cdot    {\bf w}_l =  -  { 1 \o 2 N^2};  \quad (k \ne l); \qquad   {\bf w}_k \cdot  {\bf w}_k   =   {N-1 \o 2  N^2},  \qquad
k,l=1,2,\ldots, N.
\eeq  
They are vectors in $N-1$ dimensional Euclidean space.      The quantization condition  for  a vortex  is
 \beq     U(2 \pi)  \in    {\cal C}  = \mathbb Z_N, \qquad   
U(\phi) = \exp{i \sum_j^r 
\beta_j T_j   \phi}
\label{seethis} \eeq
which reads   for the   minimum flux:
\beq   \sum_{i=1}^{N-1}  \beta^i   w_k^{i}  = {\mathbf  \beta }\cdot {\bf w}_k  =  -   { n_k \o N},   \qquad   k=1,2,\ldots  N, 
\label{suneq1} \eeq
with 
\beq  
 n_k=  \{ 1 \,\, mod  \,\, N \}, \qquad    \sum_{k=1}^{N}  
n_k=0. 
\label{suneq2}  \eeq
To find the solutions to these equations we first note that  of  $N$ equations (\ref{suneq1})   only 
$N-1$     are independent, due to the fact that 
    \beq  \sum_{k=1}^{N}   {\bf  w}_k^{i} =0. \eeq 
One particular solution of     $N$ equations (\ref{suneq1}),(\ref{suneq2})    can then be found by e.g. choosing 
\beq n_2=n_3=\ldots = n_N= 1, \quad   n_1= -  (N-1) 
\label{suneqpart}   \eeq  
With this choice,   the vector  ${\tilde \beta}=   \beta/2N$   forms   the same scalar products with  $N-1$ weight vectors ${\bf 
w}_{2},$   ${\bf  w}_{3},$  ... ${\bf  w}_{N},$  as does ${\bf  w}_{1}$.    (See Eq.(\ref{seethis}.)   This proves that
\beq      {\tilde \beta} = {\bf  w}_{1}    \qquad .^..  \quad    {\bf \beta} =    2 N  {\bf w}_1 \equiv    {\bf \beta}_1 ,  
\eeq
for if $ {\tilde \beta} - {\bf  w}_{1} $  were not a  null vector,    it would be   orthogonal to each vector in the complete set,  \{ ${\bf 
w}_{2},$   ${\bf  w}_{3},$  ... ${\bf  w}_{N}$ \},    which is an absurdity.

Other solutions of  (\ref{suneq1}), (\ref{suneq2})    can be found by choosing   different set of $N-1$  $n_i$'s to be equal to
$1$, and the remaining one (say,  $n_j$) to be equal to $-(N-1)$.  The corresponding vortex has
\beq    {\bf \beta}_j =     2 N  {\bf w}_j, \qquad j=2,3,
\ldots, N.      \label{betasol}   \eeq
We thus find that the possible vortices  of minimum flux are  characterized  precisely by the weight vectors of the
fundamental representation of $SU(N)$, generalizing the result     found for $SU(3)$. 

Note that these $N$   solutions are not independent:   they are related by  gauge    transformations hence physically
equivalent.     There is thus a unique  (minimum)
vortex of $N$-ality   one. 

The vortices of the minimum flux minus one ($N$-ality   $= -1$)  can be found analogously,  with a plus sign on the right
hand side in E.(\ref{suneq1}), hence by changing the sign of  $\beta_j$ in the solutions found above.

There are other solutions of  (\ref{suneq1}), (\ref{suneq2})  representing   vortices of higher $N$-alities.     At  the $N$-ality two, for
instance,   the  solutions for $\beta$ have   the form, 
\beq      2 N ({ {\bf  w}}_i  + { {\bf  w}}_j), \qquad    i,j = 1,2, \ldots, N.   
\eeq 
They fall into  two gauge inequivalent sets of  vortices:   their sources would    carry  the quantum numbers of the two
irreducible representations,   
\beq      \Young {2 } \,,  \quad  \Young [-1] { 1 1 }\, , 
\eeq   
symmetric and antisymmetric in color, respectively.

Solutions of   $N$-ality   $k$   can be analogously be constructed  by taking as $\beta$ the vector sum of  arbitrary $k$ 
minimum solutions,   Eq.(\ref{betasol}).    These vortices can be grouped  into gauge invariant subsets, each of which has a source 
  carrying      quantum numbers 
of an      irreducible  representations of $SU(N)$ group,  
\beq    \overbrace{\Young {3 } \ldots  \Young {2 }}^k, \qquad   \overbrace{\Young [-1]{2  1} \ldots   \Young {2 }}^{k-1},\qquad 
\ldots,   \qquad  \Young[-4]{1 1 1 1 1} \, , 
\label{allthe} \eeq   
all having $k$ boxes. 

The vortices of $N$-ality, $1,2,\ldots, N-1  $   cannot be  unwound   by a gauge transformation,   representing elements of  the
 fundamental group
\beq     \Pi_1 (SU(N)/\mathbb Z_N) = \mathbb Z_N. 
\eeq
    Nevertheless,   this does not mean that each of  the   vortices (\ref{allthe})    is  stable against decay.    A  vortex of a given $N$-ality can decay through the 
pair production  of gauge bosons   into one  of the same  $\mathbb Z_N $  quantum number but 
 with a lower   tension, via processes   
similar to the one in the $SU(3)$  example of   Fig. \ref{vortex}.         It is possible that   the tension 
is  smallest in the case of  the antisymmetric
representation
$\underline { N   \choose  k}.$      In other words,  the solution for the vortex charge $\beta$     at $N$-ality
$k$ is truely  a unique gauge-invariant set
\beq   2 N \,   \{  { {\bf  w}}_{i_1}   + { {\bf  w}}_{i_2}  + \ldots + { {\bf  w}}_{i_k} \quad  mod    \quad    \alpha \},   \qquad    i_m = 1,2, \ldots,
N,   
\eeq 
where $\alpha$'s are the root vectors of the $SU(N)$ group.

  Which of these, apart from the smallest,  $N$-ality one  vortex,  is stable against decay into   a  bundle of 
vortices with smaller   $N$-alities,    is again  a  dynamical
question   (i.e., depends on  the form of the potential, values of  coupling constants,  quantum corrections, etc.).  
    One would expect   no  universal      formula for the relative tensions
among vortices of different $N$-alities,      on the general ground.  However, 
  there are  some intriguing    suggestions \cite{STR}  
that  the ratios  among the vortex tensions  for  different $\mathbb  Z_N$ charges,   found originally    in the pure ${\cal N}=2$
supersymmetric Yang-Mills theory  (broken softly to ${\cal N}=1$) \cite{DS}, 
\beq      T_k \propto     \sin { \pi   k \o  N}, 
\eeq
might be  universal.     
The  results  from lattice
calculations   with $SU(5)$ and $SU(6)$  Yang-Mills theories 
\cite{LT,Pisa}  are consistent with the sine formula.  More recent results on these ratios  \cite{KI,KA} however 
seem  to  indicate  the non-universality of these ratios,  though the  deviation from it may be small.

The absence of   vortices  of $N$-ality, $N$,  can be understood since  the  charges   corresponding to an irreducible representation
with $N$ boxes  in the Young tableau,   can always be    screened by those of the dynamical fields (adjoint representation):
 the
vortex is broken by copious production of massless gluons   of the dual $SU(N)$  theory.

It is also  easy to prove that these solutions do satisfy the general condition  Eq.(\ref{general}).  In fact,  for each  root vector
$\alpha$,  one finds from
Eq.(\ref{betasol})  
\beq  \alpha \cdot \beta_j=  2N   \alpha \cdot {\bf w}_j  =       N   (\alpha \cdot   \alpha )  \times  {\hbox{\rm integer}},  
\eeq
where we have used the well known theorem stating that for any root vector  $  \alpha$  and for any weight vector ${\bf w}$,
   $2 (\alpha \cdot {\bf w})/ (\alpha \cdot   \alpha ) $   is an integer.     For $SU(N)$,  $ \alpha \cdot   \alpha = 1/N$, thus 
the general condition Eq.(\ref{general}) is indeed met strictly   by   our  solutions.

\subsection {$\mathbf { SO(2N)}$}         
 The  generators in the Cartan subalgebra  of $SO(2N)$  group  can be taken as 
\beq 
   T_i =    \pmatrix{ -i  w_1^i  \pmatrix{ &  1 \cr -1 &} &  & & \cr
   & - i  w_2^i\pmatrix{ & 1 \cr -1  &} &  & \cr  & & \ddots  & \cr
    &  & & -i  w_N^i\pmatrix{ & 1 \cr -1 &}}, 
\label{songene} \eeq 
($ i=1,2\ldots, N$)       where  ${\bf w}_k$ ($k=1,2, \ldots, N$)  are the weight vectors of the fundamental representation, 
living in  an $N$-dimensional Euclidean space and   satisfying
\beq  {\bf w}_k \cdot    {\bf w}_l =0;  \quad k \ne l; \qquad     {\bf w}_k \cdot  {\bf w}_k   =   {1  \o 4(N-1)}: \label{weighso}\eeq
they form a complete set of orthogonal  vectors.  
 
The form of the minimal vortex depends on the  field content of the theory.   
If   the scalar fields involved  are all   assumed to 
transform as    tensor representations of even ranks  (for instance, the antisymmetric representation), then the gauge  
group is  effectively   $SO(2N)/\mathbb Z_2$,   with
\beq   \Pi_1(SO(2N)/\mathbb Z_2) = \mathbb Z_2 \times \mathbb Z_2.  
\label{z2z2}\eeq 
 The quantization condition  for
a minimum  nontrivial
$SO(2N)$   vortex is that  $U(\phi)$   appearing in the asymptotic behavior of the fields Eq.(\ref{standard})   behaves in this case 
as
 \beq   
U(\phi) = \exp{i   \, \phi \sum_{j=1}^N  
\beta^j T_j  }   \stackrel{\phi=2\pi}    {\longrightarrow}   \pmatrix{  -{\bf 1}_{2\times 2}  &  & & \cr
   &  -{\bf 1}_{2\times 2}   &  & \cr  & & \ddots  & \cr
    &  & & -{\bf 1}_{2\times 2}}=   -{\bf 1}_{2N \times 2N}.       \label{quantison}      \eeq     
It means that  $\beta$  has   a   general form, 
\beq   { \beta}  =2 (N-1)   \, \{ \pm {\bf w}_1   \pm {\bf w}_2  \ldots    \pm {\bf w}_N \},  \label{minimum}  \eeq 
so that    
\beq    \beta \cdot   {\bf w}_i=   \pm {1 \o 2}, \qquad   i=1,2,\ldots, N.  
\eeq
To see  the consistency with the general condition, Eq.(\ref{general}),    we note    that the root vectors     of $SO(2N)$ 
group    are 
$ \alpha=\pm  {\bf w}_i \pm {\bf w}_j$ $\, (i\ne j)$.     One finds  then  
\beq    \beta \cdot \alpha = \pm 1, 0
\eeq
for any solution of  the form (\ref{minimum}).

There are $2^N$ solutions with   the minimum flux, (\ref{minimum}).  Half of them 
       ($2^{N-1}$)   contain   even number of minus  signs, the other half  an odd number of 
minus signs.      Note that   the  Weyl transformations (Eq.(\ref{weyltr}))     transform among these solutions   by permutations, 
leaving  however the set with even (or odd) number of minus signs invariant.\footnote{ For instance
   a $\pi$  rotation in the $(1-2)$
plane changes  ${\bf w}_1  \to  - {\bf w}_1,$    ${\bf w}_2  \to  - {\bf w}_2,$   while leaving other  $ {\bf w}_i$'s untouched.      }   Thus the 
$2^{N-1}$  solutions with even  number of minus signs form    an  irreducible representation.    So do 
$2^{N-1}$  solutions with odd numbers of minus signs,  in    another irreducible representation.     These are precisely the
multiplicities   of chirality $\pm 1$    spinor
representations of    
   the $SO(2N)$   group.  {\it  We   conclude that   the sources of  the    vortices   in   $SO(2N)/\mathbb Z_2 $  theory carry the quantum
numbers of  the  chirality  $\, \pm 1$ spinor representations of   the (dual)  $SO(2N)$   group.}    They represent the nontrivial elements of 
 the first homotopy group  (\ref{z2z2}).

Consider   instead  a theory in which  scalar fields in the vector (or any odd-rank tensor) representation 
get    vacuum expectation values.   The gauge group is now   truely $SO(2N)$, with 
\beq   \Pi_1(SO(2N)) = \mathbb Z_2.   
\eeq 
The vortex representing the nontrivial element of this $ \mathbb Z_2$    is of the form,  Eq.(\ref{standard}),   with 
\beq   
U(\phi) = \exp{i   \, \phi \sum_{j=1}^N  
\beta^j T_j  }   \stackrel{\phi=2\pi}    {\longrightarrow}   \pmatrix{  {\bf 1}_{2\times 2}  &  & & \cr
   &  {\bf 1}_{2\times 2}   &  & \cr  & & \ddots  & \cr
    &  & & {\bf 1}_{2\times 2}}=   {\bf 1}_{2N \times 2N} :       \label{quantison2}      \eeq   
The minimal solutions  for $\beta_i$  are then:
\beq   { \beta}  =  \pm  \,  4 (N-1)   \, {\bf w}_i,  \qquad  i= 1,2,\ldots, N: \label{minimum2}  \eeq 
corresponding to the sources in the vector (${\underline {2N}}$) representation of the (dual) $SO(2N)$
theory.

\subsection{$\mathbf {SO(2N+1)}$ }

The  $N$   generators in the Cartan subalgebra  of $SO(2N+1)$  group  can be taken as 
\beq 
   T_i =    \pmatrix{ -i  w_1^i  \pmatrix{ &  1 \cr -1 &} &  & &  &  \cr
   & - i  w_2^i\pmatrix{ & 1 \cr -1  &} &  & \cr  & & \ddots  &  &    \cr
    &  & & -i  w_N^i\pmatrix{ & 1 \cr -1 &}   &  \cr  
   &&&&  0   },  
\label{songene2} \eeq 
  where  ${\bf w}_k$ ($k=1,2, \ldots, N$)  are the weight vectors of the fundamental representation, 
living in  an $N$-dimensional Euclidean space and   satisfying
\beq  {\bf w}_k \cdot    {\bf w}_l =0;  \quad k \ne l; \qquad     {\bf w}_k \cdot  {\bf w}_k   =   {1  \o 2(2N-1)}: \label{weighso2}\eeq
they form a complete set of orthogonal  vectors.  The quantization condition   in   this case is: 
 \beq   
U(2 \pi ) = \exp{i   \, 2 \pi \sum_{j=1}^N  
\beta^j T_j  }   =    \pmatrix{  {\bf 1}_{2\times 2}  &  & &  & \cr
   &  {\bf 1}_{2\times 2}   &  &  &  \cr  & & \ddots  &  &\cr
    &  & & {\bf 1}_{2\times 2}  & \cr
&&&&& 1}=   {\bf 1}_{(2N +1) \times (2N +1) }.       \label{quantison3}      \eeq     
The minimal solutions  for $\beta_i$  are then  $2N$  solutions:
\beq   { \bf \beta}  =  \pm  \,  2 (2N-1)   \, {\bf w}_i,  \qquad  i= 1,2,\ldots, N, \label{minimum3}  \eeq 
which can be  regarded as the weight vectors of the fundamental representation of $USp(2N)$  group
which is dual to $SO(2N+1)$ (see Eq.(\ref{dualgroup})).    

As the root vectors of  $SO(2N+1)$  group  are   $\alpha=  \{\pm {\bf w}_i, \,\,   \pm {\bf w}_i \pm   {\bf w}_j \}$, 
the general condition Eq.(\ref{general})  is satisfied  minimally   (i.e.,   $\beta \cdot \alpha = \pm1$). 

On the other hand,    nonminimal solutions  obtained by combining  two minimal solutions   $\beta_1 $ and  $\beta_2$     of the form,
(\ref{minimum3}),    make up  a set   with
\beq    { { \bf \beta}\o 2(2N-1)}  =    \pm   2{\bf w}_i, \, \,    \pm {\bf w}_i \pm   {\bf w}_j: 
\label{rootusp} \eeq
these correspond (apart from an overall normalization)  precisely to the root vectors  of  the dual group    $USp(2N)$.   These
vortices   
 can be gauge-transformed away  as   $\Pi_1(SO(2N+1))= 
\mathbb Z_2$.

\subsection{$\mathbf {USp(2N)}$ }  
The  $N$     generators in the Cartan subalgebra  of $USp(2N)$  group  are the following         $2N\times 2N$ matrices,   
\beq  {\mathbf T}_i=     \pmatrix{ {\mathbf{B_i} }   &  {\mathbf{ 0 }}  \cr 
  {\mathbf{ 0 } }   &  - {\mathbf{B_i}^{t}  }  },     \quad  i=1,2,\ldots,  N, 
\eeq
where 
\beq
{\mathbf{B_i} }  =
\left( \begin{array}{ccccc}
w_{1}^{i} & & & & \\   
 & w_{2}^{i} & & & \\
 & 0 & \ddots & 0 & \\
 & & & w_{N-1}^{i} & \\  
 & & & & w_{N}^{i}
\end{array} \right) ,   \qquad         i=1,2...N.
\label{USPGEN}   \end{equation} 
The weight vectors  ${\bf w}_k$ ($k=1,2, \ldots, N$)  form a complete set of orthogonal  vectors
  in  an $N$-dimensional Euclidean space and   satisfy 
\beq  {\bf w}_k \cdot    {\bf w}_l =0;  \quad k \ne l; \qquad     {\bf w}_k \cdot  {\bf w}_k   =   {1  \o 4(N+1)}.  \label{weighusp}\eeq
   The quantization condition  for  a vortex  is  
       \beq     U(2 \pi)  =  - {\bf 1}_{2N \times 2N},   \qquad    U(\phi) = \exp{i \sum_j^r 
\beta_j {\mathbf T}_j   \phi}
\label{seethisbis} \eeq
which reads   for the   minimum flux:
\beq   \sum_{i=1}^{N}  \beta^i   w_k^{i}  = {\mathbf  \beta }\cdot {\bf w}_k  =  \pm { 1\o 2},  \qquad  k=1,2,\ldots  N.
\label{uspeq} \eeq
The minimum solutions are  
\beq   { \beta}  =2 (N+1)   \, \{ \pm {\bf w}_1   \pm {\bf w}_2  \ldots    \pm {\bf w}_N \},  \label{minimum4}  \eeq 
where signs can be  arbirarily chosen. 
These can be interpreted as   the weight vectors of the $2^{N}$  dimensional  spinor representation of   the dual group,
$SO(2N+1). $    Note that, in contrast to the case of  the $SO(2N)$ theory,   the signs  in Eq.(\ref{minimum4}) can be changed singly
by a gauge transformation.    For instance, to change the sign of  $ {\bf w}_1,$     consider an  $SU(2)$ subgroup of  $ USp(2N)$   
acting   in the
$[i,j] = [1,  N+1]$  subspace and  make a  $\pi$  rotation with  $\sigma_1/2$.    This unique gauge invariant  set of the sources
represents the element of 
\beq  \Pi_1(USp(2N)/\mathbb Z_2)=  \mathbb Z_2.  
\eeq

Also in this case the general condition  Eq.(\ref{general})   is easily seen to be fulfilled  as the root vectors of 
$USp(2N)$  are given by   (\ref{rootusp}).

\section{Non-Abelian Duality and  Confinement  in  $SU(N)$  Theories   \label{sec:duality}  }

Classification of the possible   phases of $SU(N)$ gauge theory at zero temperature  was  discussed   by 't Hooft,
 by making  crucial use of electromagnetic duality \cite{THDUAL,DW}.    The fundamental  issue  is   
  the quantum numbers of the entity  which condenses.   If magnetically  charge  particles condense
(magnetic Higgs phase),    the theory is in a confinement phase, the Wilson loop displaying the area law.   If 
electrically charged particles get VEVS, the theory is in a Higgs phase.    If no particles condense,  the theory is in a Coulomb phase,   with a
long range force between   charged particles.   Intermediate phases where particles charged both electrically and magnetically (dyons)
condense, are also possible.     In the pure
$SU(N)$  Yang Mills theory,   these charges -  the external electric charges which can be introduced as a probe and which cannot be
screened by the dynamical  fields,   and   the possible values of the magnetic charges -  are both classified   by the center
charge  of  
$SU(N)$,    $\mathbb Z_N$.   If  a field with  $(\mathbb Z_N,  \mathbb Z_N)= (a,b)$ condenses,  any particles $(c,d)$   having a nonvanishing 
relative Dirac unit with respect to it,  $ad-bc\ne   \{0 \,\, mod \,\,  N\}$,  are confined,  while those having  $ad-bc=  
\{0 \,\, mod \,\,  N\} $ are not.  

 However,  such a classification is not  quite   complete.   
In particular,  it is not clear in such an approach which the full set of effective low-energy degrees of freedom 
are  and how they interact.  It is in this respect that the results found   in \cite{CKMP}
are to be appreciated, as they  purport to answer more detailed questions about confinement.

In fact,  a   more   physical picture  of confinement   was proposed by 't Hooft  through the so-called Abelian 
gauge fixing \cite{TM}.   {\it Assuming}  that the relevant degrees  of freedom are the monopoles  that  appear as singularities of such a
gauge fixing, the confinement is understood as the dual Meissner effect of the effective 
$U(1)^2
\subset  SU(3)$ theory.       This proposal was widely studied in the literature; however, 
its correctness   remains an open question  (for instance, see \cite{DG}).

In view of the problems with  dynamical   Abelianization    pointed out  in 
\cite{STRASS,DS,YUNG},    it would be  an  important task   to assess the    plausibility, and possibility,  of  a non-Abelian variety of 
dual superconductivity as an alternative, and perhaps true,  mechanism of confinement in QCD.
Our analysis  show that the vortices appearing  in the Higgs phase of  the dual, magnetic  
 $SU(N)/\mathbb Z_N$  theory  (in the case of the original,
  $SU(N)$  Yang-Mills theory)   have  the right properties required to  confine quarks, if these   are  introduced in the 
  electric    theory.

\section*{Acknowledgments}

One of the authors (K.K.)   thanks R. Auzzi, S. Bolognesi, M. Campostrini, T. Kugo, P. Kumar,  H. Panagopoulos,  
 and E. Vicari  for stimulating   
discussions,   and
 F. A. Schaposnik,  M.A.C. Kneipp and   T. Vachaspati for bringing   the works  in \cite{DS1}-\cite{Kneipp}  to our attention.

\bigskip
\bigskip

\newpage

\appendix

\section  {Explicit solution for   $SO(3)$ vortex }

Existence of a regular vortex like solution in the $SO(3)=SU(2)/\mathbb Z_2$  theory discussed in 
Sec.~\ref{sec: so3}    can be shown by setting 
\beq    A_{\phi}^3 = A_{\phi}^3(r) \ne 0, \qquad    A_{\mu}^a =0, \quad {\hbox {\rm otherwise}}, 
\eeq
\beq   {\hat \phi_1}  =  \psi_1(r)    U_n(\phi)   \pmatrix{1\cr 0 \cr 0}  = \psi_1(r)  \pmatrix{   \cos n \phi \cr
\sin n  \phi  \cr 0} ; \label{formphi1}\eeq
   \beq
 {\hat \phi_2}  =   \psi_2(r)  U_n(\phi)  U_1(\Theta ) \pmatrix{1\cr 0 \cr 0}=    \psi_2(r)   \pmatrix{   \cos ( n \phi  + \theta) \cr
\sin  ( n \phi  + \theta) \cr 0}.  \label{ formphi1}
\eeq
The equations of motion  for $\phi_1$  are: 
\beq - \Delta  \phi_1^1-  2   g      A_{\mu}^3
 \de_{\mu}  \phi_1^2      - g^2   (A_{\mu}^3)^2 \phi_1^1  +   2 \lambda_1 ( \phi_1^a \cdot \phi_1^a -  F_1^2)  \phi_1^1 + 2 \kappa 
( \phi_1^a \cdot \phi_2^a - G^2) \phi_2^1
=0,   \label{eqphi11}
\eeq
\beq - \Delta   \phi_1^2+  2   g      A_{\mu}^3
 \de_{\mu}  \phi_1^1      - g^2   (A_{\mu}^3)^2 \phi_1^2  +   2 \lambda_1 ( \phi_1^a \cdot \phi_1^a -  F_1^2)  \phi_2^2+ 2 \kappa 
( \phi_1^a \cdot \phi_2^a - G^2) \phi_2^2
=0,     \label{eqphi21}
\eeq
and analogous ones with  $ \Delta  \phi_2^a$. 
Combining these     one    finds 
 \bea  &&  - [\, {1 \o r} {\de \o \de r} r  {\de \o \de r}  +   { 1 \o r^2}  { \de^2 \o \de \phi^2} \,]  \,  ( \psi_1  e^{i n \phi} )  +  2  \,  g 
\, i      A_{\phi }^3
 { 1 \o r} \de_{\phi }   ( \psi_1  e^{i n \phi} )    +  g^2   (A_{\phi}^3)^2  \,  ( \psi_1  e^{i n \phi} ) \non \\
&+&      2\, \lambda_1 ( \phi_1^a \cdot \phi_1^a -  F_1^2)  ( \psi_1  e^{i n \phi} )     + 2 \, \kappa 
( \phi_1^a \cdot \phi_2^a - G^2)  ( \psi_2  e^{i n \phi  + i \theta} ) 
=0,   
\eea 
We now  choose 
\beq   \theta= {\pi \o 2}; \qquad   G=0   \label{Ansatz}
\eeq
so that 
\beq  -  {1 \o r} {\de \o \de r} r  {\de \o \de r}\psi_1  +   \left({ n  \o r  }  -  g    {A_{\phi }^3  } \right)^2    \psi_1   
+         2 \lambda_1 (  \psi_1^2   -  F_1^2)     \psi_1    
=0,   
\eeq
The equation for   $\phi_2$  takes   the same form,  with  $\{\psi_1, \lambda_1, F_1\} \to  \{\psi_2, \lambda_2, F_2\}.$
These equations have the same form as in  the   $U(1)$ theory \cite{NO}.
The equation for the gauge  field is:
\beq    {1 \o r} {\de \o \de r} r  {\de \o \de r}  A_{\phi}^3  -  {  A_{\phi}^3 \o r^2}  +    ( { g  n \o r}  -    g^2   A_{\phi}^3)  
(\psi_1^2 + \psi_2^2)  =0.   \label{explain}
\eeq 
To simplify further,  let us   take  $ \lambda_1=\lambda_2$,   $F_1=F_2$,  so  $\psi_1= \psi_2\equiv \psi(r)$,  and one has  coupled 
equations
\beq  -  {1 \o r} {d \o d r} r  {d\o d r}\psi  +  {  (n    -  g   A(r))^2  \o r^2}     \psi  
+         2 \lambda(  \psi^2   -  F^2)     \psi    
=0,   
\label{quadeqn1}  \eeq
\beq    {d\o d r}  {1 \o r}  { d \o d r}  A(r)  +   2 g  {   n  -   g   A(r)  \o r}   \psi^2   =0,
\label{quadeqn2}\eeq 
where we set    $A_{\phi}^3(r) \equiv {A(r) \o r}.   $
These can be solved easily numerically. 
If the relation  
\beq   2 \lambda  = g^2
\eeq
is satisfied,     the above equations reduce further    to the linear ones, 
\beq    { d \o dr }  \psi    =      {   n   -   g   A(r) \o r}   \psi; 
\qquad      {1 \o r}  { d \o d r}  A(r)    =  - g (\psi^2 - F^2).   
\label{linear}  \eeq
as can be easily verified.
One can further  restrict oneself    to  the case of the  minimum  vortex
 with $n=1, \, $   
since  $n=2$   vortices can be gauged away (see  Appendix  B).    The  profile of this vortex    looks
very similar to the  $U(1)$  vortex of Nielsen-Olesen.

\section  {Gauge transformation which unwinds the $SO(3)= SU(2)/\mathbb Z_2  $ ``vortex"  with flux  $n=2$}

The fact that   the ``vortex" of winding two does not represent a true vortex,   follows from the group property,
$\Pi_1(SO(3))=\mathbb Z_2$.    However,  in this case it is easy to construct explicitly,  borrowing the idea from  \cite{WUYANG},    a gauge
transformation to ``unwind"  the apparent vortex-like configuration  Eq.(\ref{candidate})    with $n=2$.    

Define 
\beq    U_n(\phi) =  \pmatrix {\cos  n  \phi  &  -  \sin n  \phi    & 0   \cr     \sin n \phi  & \cos n \phi   & 0 \cr 0 & 0 & 1}, 
\eeq
\beq
\xi =  \pmatrix{ \cos (\frac{\pi }{2 + 2\,r})\,{\cos^2  \phi }+ {\sin^2 \phi }  &
   \left\{ -1 + \cos (\frac{\pi }{2 + 2\,r}) \right\}  \,\cos \phi \,\sin \phi  &
   - \cos \phi \,\sin (\frac{\pi }{2 + 2\,r})   \cr
 \left\{ -1 + \cos (\frac{\pi }{2 + 2\,r}) \right\} \,\cos \phi \,\sin \phi  & 
   {\cos^2 \phi }+ \cos (\frac{\pi }{2 + 2\,r})\,{\sin ^2\phi }   &  
   -\sin (\frac{\pi }{2 + 2\,r})\,\sin \phi  \cr
    \cos \phi \,\sin (\frac{\pi }{2 + 2\,r})  &   \sin (\frac{\pi }{2 + 2\,r})\,
  \sin \phi   &  \cos (\frac{\pi }{2 + 2\,r})},    
  \eeq
and 
\smallskip
\beq
\eta = \pmatrix  { a &  b & c \cr  d  & e  & f \cr  g & h &  k },  
\eeq
where    
\bea   a&=&   -\cos \phi\,\sin (\frac{\pi }{1 + r})\,\sin (\frac{\pi }{2 + 2\,r})    + 
    \cos (\frac{\pi }{1 + r})\!\left\{ \cos (\frac{\pi }{2 + 2\,r})\,\cos^2 \!\phi+ {\sin^2  \!\phi} \right\},   \non \\
   b &=&   \left\{ -1 + \cos (\frac{\pi }{2 + 2\,r}) \right\} \! \cos \phi\,\sin \phi,  \non \\
c&=&   -\cos (\frac{\pi }{1 + r}) \cos \phi  \sin (\frac{\pi }{2 + 2\,r})   - 
    \sin (\frac{\pi }{1 + r})\!\left\{ \cos (\frac{\pi }{2 + 2\,r})  \cos^2 \!\phi + {\sin^2  \!\phi}  \right\}, \non \\
 d&=&  -\left\{  \sin (\frac{\pi }{1 + r})\,\sin (\frac{\pi }{2 + 2\,r})  + 
      2\,\cos (\frac{\pi }{1 + r})\,\cos \phi  \,\,{\sin^2 (\frac{\pi }{4 + 4\,r})}  \right\}     \sin \phi, \non \\
 e&=& \cos^2 \!\phi + \cos (\frac{\pi }{2 + 2\,r})\,{\sin^2  \!\phi}, \non \\
f&=&    - \cos (\frac{\pi }{1 + r})\,\sin (\frac{\pi }{2 + 2\,r})\,\sin \phi     + 
    \sin (\frac{\pi }{1 + r})\,  {\sin^2  (\frac{\pi }{4 + 4\,r})}\,\sin (2\,\phi ), \non \\
g&=&  \cos (\frac{\pi }{2 + 2\,r})\,\sin (\frac{\pi }{1 + r}) + 
    \cos (\frac{\pi }{1 + r})\,\cos \phi\,\sin (\frac{\pi }{2 + 2\,r}), \non \\
h&=&  \sin (\frac{\pi }{2 + 2\,r})\,\sin \phi, \non \\
k&=&  \cos (\frac{\pi }{1 + r})\,\cos (\frac{\pi }{2 + 2\,r}) - 
    \cos \phi\,\sin (\frac{\pi }{1 + r})\,\sin (\frac{\pi }{2 + 2\,r}),  \non \\
\eea
and consider   a gauge transformation,
\beq    {\hat \phi_A}    \to   U_{global}( r, \phi)    {\hat \phi_A} ; \qquad  
A_i  \to    U_{global}( r, \phi)  (A_i    +  { 1 \o g}  {\de}_i )   \,   U_{global}^T ( r, \phi),   
\eeq
where   
\beq   U_{global}( r, \phi) \equiv   \eta(\phi, r) \cdot  U_2( -\phi) \cdot \xi(\phi, r).  
\eeq
It can be explicitly checked that   $ U_{global}( r, \phi) $   is regular everywhere  and   that 
\beq    U_{global}( r, \phi)  \stackrel { r \to \infty} {\longrightarrow } U_2( -\phi); 
\qquad   U_{global}( r, \phi)   \stackrel { r \to 0  } {\longrightarrow } {\bf 1},
\eeq
so that the ``flux"  disappears    in the new gauge.

\end{document}